\documentclass[11pt]{article}
\usepackage[a4paper,bindingoffset=0.in,left=2.5cm,right=2.5cm,top=2cm,bottom=3cm]{geometry}
\usepackage{amsmath,amssymb,amsthm,amsfonts}
\usepackage{graphicx}
\usepackage{color}
\usepackage{url}
\usepackage[utf8]{inputenc}
\usepackage{lmodern}
\usepackage{setspace}
\usepackage{wrapfig}
\usepackage{mdframed}
\usepackage{subcaption}
\usepackage{algorithm}
\usepackage[a]{esvect}
\usepackage[noend]{algorithmic}
\usepackage{tikz}
\usepackage{enumitem}
\usetikzlibrary{tikzmark}

\begin{document}

\title{Counting Causal Paths in Big Times Series Data on Networks}

\author{Luka V. Petrovi\'c$^1$ and Ingo Scholtes$^2$ \medskip \\
\small Data Analytics Group \\ 
\small Department of Informatics (IfI) \\ 
\small University of Zurich \\
\small Z\"urich, Switzerland \\
\small $^1$\texttt{petrovic@ifi.uzh.ch}, $^2$\texttt{scholtes@ifi.uzh.ch}}

\date{}
\maketitle



\begin{abstract}
  Graph or network representations are an important foundation for data mining and machine learning tasks in relational data.
  Many tools of network analysis, like centrality measures, information ranking, or cluster detection rest on the assumption that links capture direct influence, and that paths represent possible indirect influence.
  This assumption is invalidated in time-stamped network data capturing, e.g., dynamic social networks, biological sequences or financial transactions.
  In such data, for two time-stamped links (A,B) and (B,C) the chronological ordering and timing determines whether a causal path from node A via B to C exists. 
  A number of works has shown that for that reason network analysis cannot be directly applied to time-stamped network data.
  Existing methods to address this issue require statistics on causal paths, which is computationally challenging for big data sets.

  Addressing this problem, we develop an efficient algorithm to count causal paths in time-stamped network data.
  Applying it to empirical data, we show that our method is more efficient than a baseline method implemented in an OpenSource data analytics package.
  Our method works efficiently for different values of the maximum time difference between consecutive links of a causal path and supports streaming scenarios.
  With it, we are closing a gap that hinders an efficient analysis of big time series data on complex networks.
\end{abstract}

\maketitle

\section{Introduction}

Graph or network models are an important foundation for data mining and machine learning in relational data.
Social network analysis, node ranking, graph mining, and node embedding techniques help us to identify influential actors in social networks, find cluster or community structures in graphs, or detect nodes that exhibit anomalous behaviour.
Many of those techniques are grounded in the notion of \emph{paths} in a network.
PageRank~\cite{page1999PageRank}, e.g., computes visitation probabilities of a surfer following random paths across hyperlinks.
The clustering algorithm InfoMap \cite{rosvall2008maps} compresses paths generated by a random walker to detect community structures in graphs.
Similarly, node embedding algorithms like node2vec \cite{grover2016node2vec} use random paths to embed nodes into low-dimensional vector spaces.

A fundamental assumption behind those techniques is that links capture direct influence, while paths capture the possible \emph{indirect} influence between nodes.
However, a growing amount of data not only captures direct node interactions but also at which specific times those interactions occur.
Examples include time-stamped data on user interactions in social media, temporal neuronal activities, time-stamped financial transactions, or temporal sequences of flights between airports.
This time information complicates the mining of \emph{temporal network} data~\cite{holme2015modern}.
A major source of complexity is that transitive paths that are assumed to exist in network models might actually be infeasible in the temporal interaction sequence.
As an example, if Alice communicates with Bob and Bob communicates with Carol, a rumour can only spread from Alice via Bob to Carol, if the Alice interacted with Bob \emph{before} Bob interacted with Carol.
If the chronological order of interactions is reversed, a transitive \emph{time-respecting} or \emph{causal path} from Alice via Bob to Carol does not exist.
Hence, temporal information can invalidate the implicit assumption of static network models that paths are \emph{transitive}, thus questioning the application of graph mining and network analysis techniques~\cite{holme2015modern,Lambiotte2019_Understanding}.

To address this challenge, a growing number of works utilises \emph{higher-order network models}, which can be used to generalise network analysis and graph mining techniques to time series data~\cite{scholtes2014causality,Rosvall2014_Memory,xu2016representing}.
The basic idea behind such higher-order models is, rather than exclusively focusing on direct interactions, to additionally model the causal paths by which nodes can \emph{indirectly} influence each other~\cite{Lambiotte2019_Understanding}.
The inference of higher-order models requires \emph{statistics of causal paths}, which 
enables us to reason about the causal topology, i.e. who can influence whom, in temporal networks.

The application of this approach is currently hindered by a lack of scalable algorithms to efficiently count causal paths in big time series data.
Addressing this gap, the contributions of our work are:
\begin{itemize}
  \item We introduce and formalize the problem of counting causal paths in temporal networks, which generalises the counting of time-stamped links commonly used to construct \emph{weighted links} in time-aggregated network models.
  \item We propose the streaming algorithm to count causal paths up to a given maximum length in big time series data on temporal networks. We theoretically and experimentally show that the computational complexity of our algorithm linearly scales with the number of time-stamped links in time series data. 
  \item We show that our algorithm works efficiently for different maximum time differences between consecutive links that allow to tailor the definition of causal paths to the time scale of data.
  \item We demonstrate our method in real-world data and show that it outperforms a baseline algorithm implemented in an OpenSource data analytics package for temporal network analysis.
\end{itemize}

\section{Problem and Background}
\label{Sec:Preliminaries}

We first specify the type of time-stamped network data and define the notion of \emph{causal paths} underlying our work. We then formally define the problem addressed by our algorithm.

\begin{figure}[!ht]
  \centering
  \includegraphics[width=.8\textwidth]{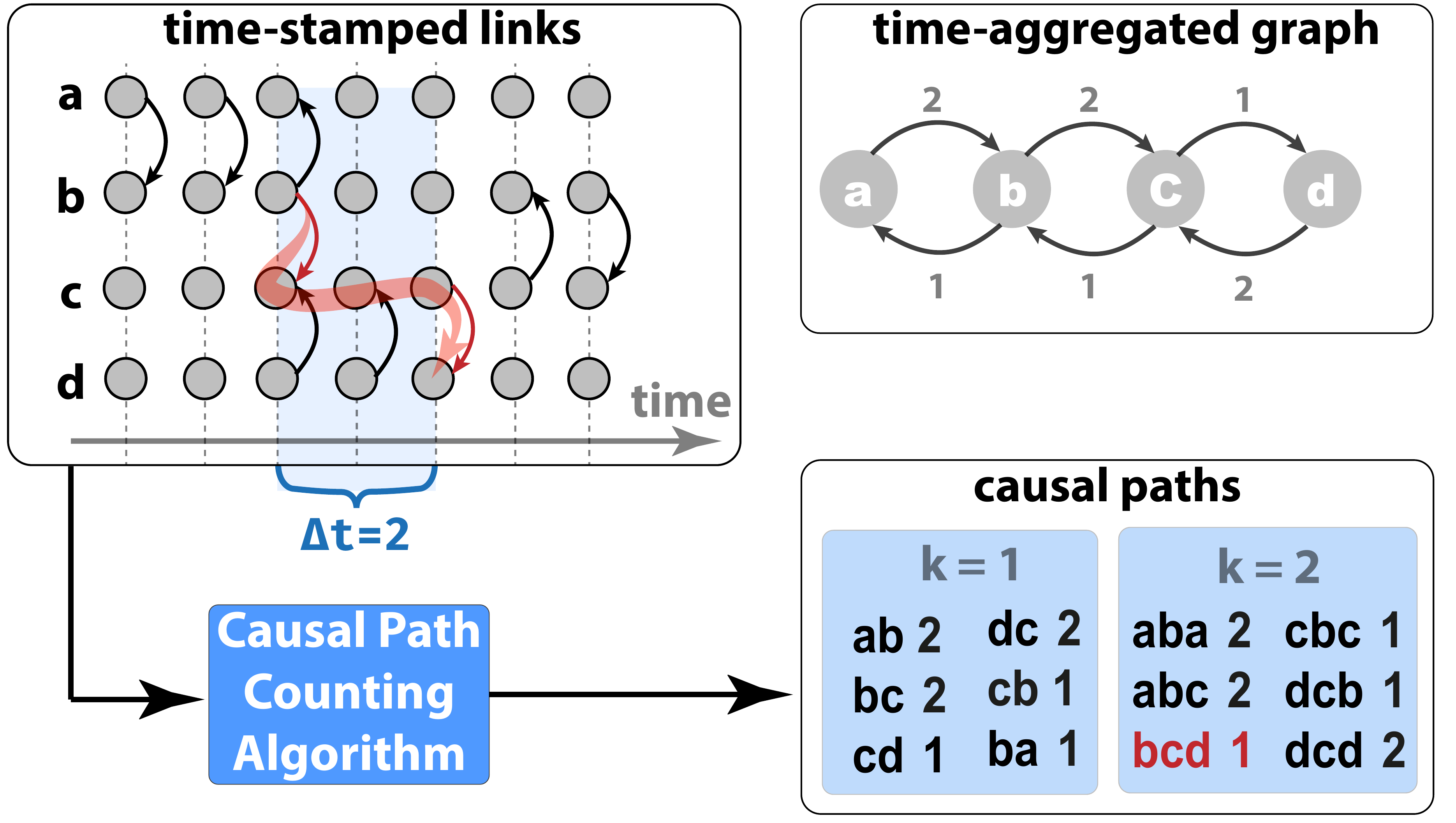}
  \caption{For a set of time-stamped links between nodes in a graph, our algorithm efficiently counts \emph{causal paths} up to a given length $K$ (right) for maximum time difference $\delta$. In the example above, for $\delta=2$ the time-stamped links $(b,c,t=3)$ and $(c,d,t=5)$ contribute to a causal path $\protect\vv{bcd}$ of length two, of which a single instance occurs in the data.}
  \label{fig:input-output}
  \end{figure}

We consider a set $D$ of (directed) time-stamped links $(s, d, t) \in V \times V \times \mathbb{N}$ where $s, d \in V$ are the source and target node of the link respectively, and $t \in \mathbb{N}$ is the discrete time stamp of an instantaneously occurring link $(s,d)$ (see left panel in Fig. \ref{fig:input-output}).
We call a sequence 
\[p:=\vv{n_0 n_1\ldots n_l}\] 
of nodes $n_i\in V$ a \emph{causal path} for a maximum time difference $\delta \in \mathbb{N}$ iff  $\forall i \in \{0, \ldots, l-1\}$
\begin{enumerate}
  \item $e_i:=(n_i, n_{i+1}, t_i) \in D$, 
  \item $t_{i+1}>t_i$, and 
  \item $t_{i+1}-t_i\leq \delta$ 
\end{enumerate}
We highlight that the second condition in the definition above forces sequences of time-stamped links on a causal path to respect the chronological order.
Ignoring this restriction would lead to a definition of paths by which nodes can influence each other backwards in time.
For any $\delta<\infty$, the third condition limits the maximum time difference between any two consecutive links on a causal path.
Clearly, different systems require different values for $\delta$, e.g. for the propagation of rumours in social networks the value of $\delta$ is linked to the ``memory'' of actors, while for the propagation of an epidemic it is linked to the time to recovery from an infectious disease.
The detection of the optimal time scale $\delta$ to analyse a given data set is an important problem by itself \cite{caceres2013temporal} that we do not address, i.e. for the purpose of our work, we treat the parameter $\delta$ as given.

We further define the length $\|p\|$ of path $p=\vv{n_0 n_1 \ldots n_l}$ as the number $l$ of traversed links.
Moreover, we call any sequence of time-stamped links $(e_0, \ldots, e_{l-1})$ that constitute a causal path $p$ one \emph{instance of the causal path $p$} in $D$.
With this, we formulate the problem addressed by our algorithm:
Given a data set $D$ of time-stamped links, a maximum time difference $\delta$, and $K\in\mathbb{N}$ we want to calculate the number of instances $C(p)$ of all causal paths $p$ with $\|p\|\leq K$.
Formally, we define
\begin{align*}
  C(p) = \left|\{ \left(e_1, \ldots, e_k\right) | \left(e_1, \ldots, e_k\right) \text{ is an instance of } p \text{ in } D \}\right|
\end{align*}
A toy example for a set of time-stamped links $D$ and the resulting output $C(p)$ for $\delta=2$ and $K=2$ is shown in Fig.~\ref{fig:input-output}.
For illustrative purposes, in the left panel of Fig.~\ref{fig:input-output} we show a so-called \emph{time-unfolded} representation of a temporal network, where each node is represented by multiple \emph{temporal copies}.
We note that for a maximum time difference of $\delta=2$ time units, this example contains two instances $((a,b,1), (b,c,3))$ and $((a,b,2),(b,c,3))$ of the causal path $\vv{abc}$ and thus $C(\vv{abc})=2$.
This captures the fact that, e.g., an epidemic can spread in two different ways from node $a$ via node $b$ to node $c$.

The rationale behind our definition of $C(p)$ -- and thus the motivation of our work -- is that we want to count in how many different ways nodes can (indirectly) influence each other in a temporal network.
It provides a natural generalisation of \emph{weighted links}, which count instances of time-stamped links in time-aggregated network models~\cite{holme2015modern}.
Since each time-stamped link is an instance of a causal path with length one, $C(p)$ extends the notion of link weights to causal paths of arbitrary length.
This is an important basis to generalise static network models to higher-order models~\cite{Lambiotte2019_Understanding}.

\paragraph{Related Work}
Approaches to find patterns in temporal networks that are related to our work exist in several research areas, namely (i) works using time-respecting paths in temporal network analysis, (ii) approaches calculating or estimating the statistics of paths as a basis for higher-order network models, and (iii) methods to efficiently analyse so-called \emph{network motifs} in temporal data.

Algorithms to calculate causal or time-respecting paths in area (i) have generally focused on generalisations of path problems in graph theory to temporal networks~\cite{Kempe2000}.
As such, most works in this line of research have considered algorithms to compute shortest or fastest causal paths~\cite{wu2014path}, the accessibility of nodes via causal paths with at most length $k$~\cite{Whitbeck2012,Lentz2013}, or the generalisation of path-based centralities to causal paths in temporal networks~\cite{Takaguchi2016_CoverageCentrality}.

While the works in area (ii) are an important motivation for our algorithm they have not focused on algorithmic aspects of efficiently counting causal paths.
They have thus either used heuristic approaches to generate causal path statistics based on simple stochastic models~\cite{salnikov2016using} or adopted simple algorithms that do not scale to large data sets~\cite{scholtes2017network}.

Taking a different perspective, counting paths in a temporal network can be considered a special case of counting temporal motifs~\cite{paranjape2017motifs}.
We highlight that, while the causal paths counted by our method for a given $K$ and $\delta$ are necessarily $K$-edge $K\cdot\delta$ temporal motifs as defined, e.g., in~\cite{paranjape2017motifs}, the opposite is not true.
Hence, our work focuses on a computational problem that, while being less general then the problem of counting arbitrary temporal motifs, is nevertheless of crucial importance to study \emph{causal topologies} in temporal networks.
To the best of our knowledge, our work is the first to provide a fast streaming algorithm to count causal paths in temporal networks for given maximum time difference $\delta$ and maximum path length $K$.

\section{Counting Causal Paths}
\label{sec:algorithm}

We now introduce an efficient streaming algorithm, which counts all instances of causal paths of lengths $k\leq K$ for a given maximum path length $K$ and a maximum time difference $\delta$ between subsequent links on causal paths (cf. section \ref{Sec:Preliminaries}).
It uses an iterative approach that incrementally extends causal paths by moving a sliding time window of length $\delta$ over the data.
Initially, each time-stamped link $(n_1, n_2, t)$ is a causal path $\vv{n_1 n_2}$ of length one that can be extended by time-stamped links in time window $\left(t, t+\delta\right]$.
To denote the extension of a causal path, we define a binary operator $\oplus$ on a path $\vv{n_0\ldots n_k}$ and a node $n_{k+1}$ as follows:
  \[ \vv{n_0\ldots n_k} \oplus n_{k+1} := \vv{n_0\ldots n_kn_{k+1}} \]

We assume that the time-stamped links in the data $D$ are chronologically ordered, where $e_i$ refers to the $i$-th link in the ordered sequence.
Our proposed algorithm performs a single pass through this sequence and, for each time-stamped link $e_i$, computes the count $C_i$ of causal path instances ending with link $e_i$. 
We define \(C_i (p) := m \), where $m$ counts different instances of causal path $p$ that end with $e_i$.

To calculate the output $C$, we iteratively aggregate path counters $C_i$ by a simple addition, i.e. we sum two path counters $C$ and $C'$:
$$C_{sum} = C + C' \iff  \forall p: C_{sum}(p) = C(p) + C'(p) $$
We denote the operation $ext_K(C,n)$ that takes a path counter $C$ and node $n$ and extends each path in $C$ with $n$, thus obtaining $C_{ext}$:
$$C_{ext} = ext_K(C, n) \iff \forall p,  \|p\|<K: C_{ext}(p\oplus n)=C(p)$$

Our method can now be formulated as shown in Algorithm~\ref{alg:paco}.
\begin{algorithm}\small
\caption{Calculate occurrences of all causal paths in data $D$ that are at most $K$ steps long, assuming parameter $\delta$}\label{alg:paco}
\begin{algorithmic}[1]
\REQUIRE $D$, $\delta$, $K$
\STATE $c \gets $ dict()
\STATE $W \gets $ list()
\FOR {$i$ in $1$ to $N$}
  \STATE $(s,d,t) \gets D[i]$
  \STATE $c_i \gets$ dict()
  \STATE  $c_i [\vv{s d}] \gets  1$
  \FOR {$j$ in $1$ to $W.length()$}
    \STATE $(s_j, d_j, t_j, c_j) \gets W[j]$
    \IF {$t_j < t -\delta$}
      \STATE $W \gets W.remove((s_j, d_j, t_j, c_j))$
    \ELSE
      \IF {$d_j=s \wedge t>t_j$}
        \FOR {$p$ in $c_j.keys()$ }
          \IF {$ \|p\| <K$ }
            \IF {$p\oplus d$ not in $c_i.keys()$}
              \STATE $c_i[p\oplus d] \gets c_j[p]$
            \ELSE
              \STATE $c_i[p\oplus d] \gets c_i[p\oplus d] +c_j[p]$
            \ENDIF
          \ENDIF
        \ENDFOR
      \ENDIF
    \ENDIF
  \ENDFOR
\FOR {$p$ in $c_i.keys()$}
  \IF {$p$ not in $c.keys()$}
    \STATE $c[p] \gets c_i[p]$
  \ELSE
    \STATE $c[p] \gets c[p] +c_i[p]$
  \ENDIF
\ENDFOR
\STATE $W.append((s, d, t, c_i))$
\ENDFOR
\RETURN $C$
\end{algorithmic}
\end{algorithm}
We explain it step-by-step and illustrate it in Fig.~\ref{table:data,one step}, using the toy example from Fig.~\ref{fig:input-output}.
We assume time-stamped links to be stored in a list $D$ with length $N$, while $\delta$ and $K$ are integers.
We further assume that the path counts $C_i$ as well as the overall output $C$ are implemented as dictionaries $c_i$ and $c$ respectively, which have causal paths $p$ as keys and integer counts as values.
We do not store paths whose count is zero.
In lines 1 and 2 we initialise the path count dictionary $c$ as well as the time window $W$.
In line 3 we iterate through the ordered sequence of time-stamped links $(s,d,t)$.
In each iteration we first initialize the dictionary $c_i$ that represents $C_i$ (line 5).
Since a time-stamped link $(s,d,t)$ is a causal path $\vv{s d}$ of length one we set its counter to one (line 6).
We then iterate through the time window $W$ (line 7 and 8), which contains tuples $(s_j, d_j, t_j, c_j)$ representing all time-stamped links $(s_j, d_j, t_j)$ and causal path counts $c_j$ that could be continued as a causal path by link $(s,d,t)$.
The red box in Fig.~\ref{table:data,one step} highlights the time-stamped links and path counts in time window $W$ at iteration $i=6$ for $\delta=2$.
If the time-stamped link $(s_j, d_j, t_j)$ in the time-window is outdated (lines 9 and 10), we remove both the link and its dictionary $c_j$ from $W$.
Otherwise, if a time-stamped link $(s_j, d_j, t_j)$ in $W$ is continued by $(s,d,t)$ (lines 11 and 12) each causal path $p$ that ends with $(s_j, d_j, t_j)$ is extended by node $d$. 
We thus found new causal paths $p \oplus d$ and update $c_i$ such that $C_i \gets C_i + ext_K(C_j, d)$ (lines 13-18).
We next update $c$ such that $C \gets C+C_i$ (lines 19-23) and extend the window by one time-stamped link $D[i]$ and path counts $c_i$ (line 24).
Once we have completed a single pass over the time-stamped links, the dictionary $c$ contains counts of all causal path instances with length $k \leq K$ and maximum time difference $\delta$.

\def\myFrColor{black}
\def\myScColor{red}

\begin{figure}[!htbp]
  \centering
  \renewcommand{\arraystretch}{1.4}
  \begin{tabular}[b]{cccc|l}

    \hline
    $i$ & $s_i$ & $d_i$ & $t_i$ & $c_i$\\
    \hline
    0 & a & b & 1 & $\vv{a b}:1$\\
    1 & a & b & 2 & $\vv{a b}:1$\\
    {\color{\myFrColor} 2} & \tikzmark{startup}{\color{\myFrColor} b} & {\color{\myFrColor} a} & {\color{\myFrColor} 3} & $\vv{b a}:1, \vv{a b a}:2$\\
                      {\color{\myFrColor}3} & {\color{\myFrColor}b} & {\color{\myFrColor}c} & {\color{\myFrColor}3} &  $\textcolor{\myScColor}{\vv{b c}}: 1, \vv{a b c}: 2$\\
                      {\color{\myFrColor}4} & {\color{\myFrColor}d} & {\color{\myFrColor}c} & {\color{\myFrColor}3} & $\textcolor{\myScColor}{\vv{d c}}:1$\\
                      {\color{\myFrColor}5} & {\color{\myFrColor}d} & {\color{\myFrColor}c} & {\color{\myFrColor}4} & $\textcolor{\myScColor}{\vv{d c}}:1$ \tikzmark{endup}\\
                      {\color{blue} 6} & {\color{blue} c} & {\color{blue} d} & {\color{blue} 5} & $\textcolor{blue}{\vv{c d}}:1, \vv{\textcolor{\myScColor}{b c} \textcolor{blue}{d}}: 1, \vv{\textcolor{\myScColor}{d c} \textcolor{blue}{d}} : 2 $\\
    7 & c & b & 6 & $\vv{c b} : 1, \vv{d c b}:1$\\
    8 & b & c & 7 & $\vv{b c}:1, \vv{c b c}:1 $\\
    \hline
    \begin{tikzpicture}[remember picture,overlay]
    \foreach \Val in {up}
    {
    \draw[rounded corners,red,thick]
    ([shift={(-0.5\tabcolsep,2.6ex)}]pic cs:start\Val) 
    rectangle 
    ([shift={(5.5\tabcolsep,-01ex)}]pic cs:end\Val);
    }
    \end{tikzpicture}
  \end{tabular}
  \caption{
Illustration of one iteration of our algorithm in the example shown in Fig.~\ref{fig:input-output}.
The left columns contains the data, while the right column shows path counters $c_i$ for $\delta = 2$ and maximum path length $K=2$.
The red box shows time window $W$ of links that could be continued by the considered link (iteration $i=6$, highlighted in blue).
The right column shows causal path counts stored in $c_i$ after iteration $i=6$, with paths from $W$ extended in $i=6$ shown in red.
}
  \label{table:data,one step}
\end{figure}

\paragraph{Computational Complexity}

In the following, we briefly comment on the computational complexity of our algorithm.

Assuming that the maximum number of links in the time window is $m_\delta$ for $\delta<\infty$, and that the input sequence $D$ is ordered, the computational complexity of our method is $\mathcal{O}( N |V| K^2 [ m_\delta \lambda_{\text{max}}^{K-2} + \lambda_{\text{max}}^{K} ])$, where $\lambda_{\text{max}}$ is the largest eigenvalue of the binary adjacency matrix of the time-aggregated network $G=(V,E)$.
We note that the maximum eigenvalue $\lambda_{\text{max}}$ is called \emph{algebraic connectivity} and is linked to the sparseness of a graph topology.
$\lambda_{\text{max}}$ takes a maximum value of $|V|$ for a fully connected graph $G=(V,E)$.
For the special case of $\delta\rightarrow\infty$, $m_\delta$ scales as $N$ and the scaling is at most quadratic in the size of the input sequence $D$.

For the more detailed derivation of the bounds above, we first consider that the maximum number of links that have the same time-stamp $t$ is bounded above by a constant $m$ for all $t$. 
The upper bound for the number of links in any time window is then $m_\delta = m \delta$ for $\delta<\infty$.
Assuming the number $\Lambda(K)$ is the upper bound for the possible number of causal paths with length up to $K$ computational complexity is:
\[\mathcal{O}\left( N \cdot K \cdot \left[ m_\delta \delta \Lambda(K-2) + \Lambda(K) \right]\right)\]
We note that the maximum number of (causal) paths of a given length $k$ depends on the topology of the underlying graph.
The maximum number of paths on a time-aggregated, directed graph is bounded above by the number of paths in its undirected version $G$.
We can thus derive an upper bound for $\Lambda(K)$ based on the topology of the time-aggregated, undirected, graph $G=(V,E)$ of time-stamped links in $D$.
We note that the number of paths of length $k$ in a graph $G$ can be calculated as 
\[\sum_{i,j}(\mathbf{A}^k)_{ij},\]
where $\mathbf{A}$ is the binary adjacency matrix of $G$ and $\mathbf{A}^k$ is the $k$-th matrix power of $\mathbf{A}$.
Assuming $ \mathbf{A}$ is positive semi-definite, we can decompose it to $\mathbf{A} = \mathbf{Q\Lambda Q}^T$ where $\mathbf{\Lambda}$ is a diagonal matrix, consisting of positive eigenvalues.
We can decompose $\mathbf{\Lambda}$ to $\mathbf{\Lambda} = \sqrt\mathbf{\Lambda} \sqrt\mathbf{\Lambda}$, and insert $\mathbf{Q}^T \mathbf{Q} = \mathbf{1}$ in between: 
$$\mathbf{A} = \mathbf{Q} \sqrt\mathbf{\Lambda}  \mathbf{Q}^T \mathbf{Q} \sqrt\mathbf{\Lambda}  \mathbf{Q}^T $$
We denote $\mathbf{Q} \sqrt\mathbf{\Lambda}  \mathbf{Q}^T$ as $\mathbf{S}$, therefore $\mathbf{A}= \mathbf{S}^T \mathbf{S}$.
The maximum eigenvalue of $\mathbf{S}$, $\lambda_{\text{max}}(\mathbf{S})$ is equal to the square root of $\lambda_{\text{max}}(\mathbf{A})$.
Then, by the definition of the induced norm for matrices, we have: 
$$\forall x, \|\mathbf{S}\|_2> \frac{\|\mathbf{S}x\|_2}{\|x\|_2}$$

Since $\|\mathbf{S}\|_2 = \lambda_{\text{max}}(\mathbf{S})$, $\|Sx\|_2 = \sqrt{x^T \mathbf{S}^T \mathbf{S} x}$ and $\|x\|_2= \sqrt{x^Tx}$ we have:

$$\sqrt{\lambda_{\text{max}}(\mathbf{A})} =\lambda_{\text{max}}(\mathbf{S})> \frac{\sqrt{x^T \mathbf{S}^T S x}}{\sqrt{x^Tx}}$$
In particular for $x = (1,1,1,\ldots,1)^T$, where $x$ is $n$-dimensional, and $x^Tx = n$:
$$\sqrt{\lambda_{\text{max}}(\mathbf{A})} >\frac {\sqrt{\sum_{i,j} \mathbf{S}^T \mathbf{S}}}{\sqrt{n}}  = \frac{\sqrt{\sum_{i,j} \mathbf{A}_{i,j}}}{\sqrt{n}} $$
Thus
$$n \lambda_{\text{max}}(\mathbf{A})> \sum_{i,j} \mathbf{A}_{i,j}$$

We can can thus derive an upper bound for the number of paths of lenth $k$:
\[ \sum_{i,j}\mathbf{A}^k_{ij} \leq |V| \cdot \lambda_{\text{max}}^k \]
where $\lambda_{\text{max}}$ is the largest eigenvalue of the adjacency matrix $\mathbf{A}$, which is also called the \emph{algebraic connectivity} of a graph.
For $\Lambda(K)$ we thus obtain 
\[ \Lambda(K) = \sum_{l=1}^K \sum_{i,j}\mathbf{A}^l_{ij} \leq \sum_{l=1}^K |V| \cdot \lambda_{\text{max}}^l \]
Assuming $\lambda_{max}>1$, we find the upper bound:
\[\mathcal{O}\left( N \cdot |V| \cdot K^2 \cdot \left[ m_\delta \delta \lambda_{\text{max}}^{K-2} + \lambda_{\text{max}}^{K} \right]\right)\]
We note that the worst case complexity holds for a fully connected graph, where all sequences of $k$ nodes constitute a possible causal paths of length $k$.
In this case we have $\lambda_{\text{max}}=|V|$ and the upper bound is 
\[\mathcal{O}\left( N \cdot K^2 \cdot \left[ m_\delta \cdot \delta \cdot |V|^{K-1} + |V|^{K+1} \right]\right)\]
In the extreme case of $\delta=\infty$, $m_\delta$ scales as $N$ and the complexity of our algorithm thus scales quadratically with $N$, i.e.

\[\mathcal{O}\left( N \cdot K^2 \cdot \left[ N \cdot |V|^{K-1} + |V|^{K+1} \right]\right).\]

This shows that, for any finite maximum time difference $\delta$ used in the definition of causal paths, the runtime of our algorithm scales linearly with the size of the data $N$. 
The multiplicative factor in the term grows exponentially with the maximum causal path length $K$, with the base that depends on the sparsity of the graph captured in the leading eigenvalue.
We finally highlight that in real-world settings we are typically interested in causal path statistics for small values of $K$, corresponding to the maximum order of higher-order graphical models of causal paths.

\section{Experimental Results}

We present results of preliminary experiments in which we asses the runtime of our algorithm for different sizes $N$ of data, for different maximum time differences $\delta$, and for different values of the maximum length $K$ of causal paths.
We compare the runtime of our method to a baseline algorithm, which has been used to compute causal paths in time-stamped data~\cite{scholtes2017network}.
We chose this baseline since (i) other works focus on computing \emph{shortest} causal paths, which is a different problem, and (ii) it is available in the Open Source package \texttt{pathpy}\footnote{\url{www.pathpy.net}}.
To count causal paths, the baseline algorithm performs three steps:
First, adopting an approach similar to~\cite{Kempe2000}, time-stamped links are represented as direct acyclic graph (DAG).
Second, the DAG is used to compute all causal paths between root and leaf nodes.
And third, all shorter paths of length $k$ contained in those longest causal paths are counted.
Different from the sliding window approach of our method, this algorithm is not suitable for streaming scenarios.

We assess the runtime of our algorithm in empirical data capturing time-stamped proximities between students, which was collected by the RealityMining project~\cite{Eagle2006}.
It captures $N=1086404$ time-stamped links over a period of more than six months between $|V|=96$ nodes.
Due to the density of the data, calculating causal paths in this data is challenging with the baseline algorithm, in particular as $\delta$ increases.

\begin{figure}[!ht]
  \centering
    \includegraphics[width=\textwidth]{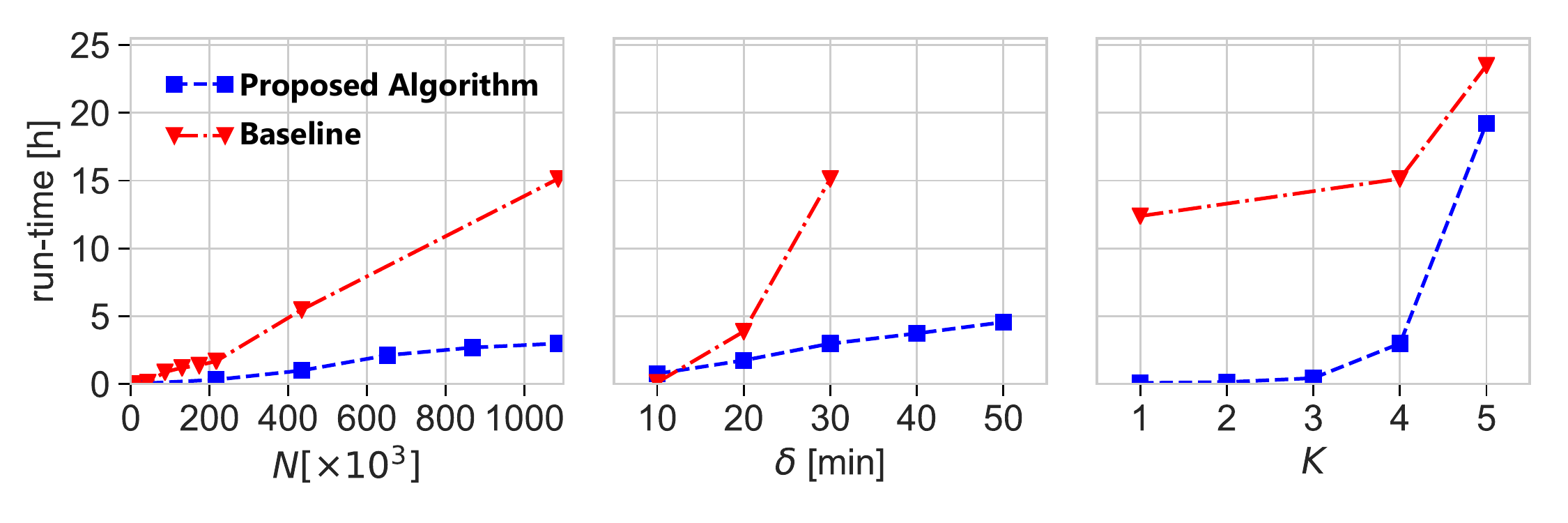}
    \caption{The left panel shows how runtime (y-axis) depends on the number of time-stamped links $N$ (x-axis) for fixed values of $\delta = 30$ minutes and $K = 4$.
    The middle panel shows how the time needed to process the whole data (y-axis) for fixed $K=3$ depends on $\delta$ (x-axis).
    The right panel shows how the runtime depends on $K$ for fixed $\delta = 30$ minutes.}
    \label{fig:results}
\end{figure}

Fig.~\ref{fig:results} shows the results of our preliminary experiments.
The left panel shows the runtime for a fixed value of $\delta$ that corresponds to 30 minutes, and a fixed maximum causal path length $K=4$.
We show the dependency of the runtime on the size $N$ of the data by running the algorithm on the first $N$ time-stamped interactions in the RealityMining data set.
The results clearly show that our approach outperforms the baseline algorithm for all values of $N$.
They further support the linear scalability of our algorithm found in section~\ref{sec:algorithm}.

The middle panel shows the dependency of the runtime on the maximum time difference $\delta$ for $K=4$.
Supporting the theoretical analysis in section~\ref{sec:algorithm}, we find that the runtime of our algorithm linearly depends on $\delta$.
Due to the calculation of causal paths between root and lead nodes in the DAG, the runtime of the baseline algorithm superlinearly grows with $\delta$, which makes it unsuitable for big data.
For the baseline algorithm we had to skip values of $\delta$ larger than 30 minutes due to the prohibitively long runtime of those experiments.

Finally, the results in the right panel show the dependency of the runtime on the maximum length $K$ of causal paths.
The result again supports our theoretical analysis, which yields a multiplicative factor that exponentially scales in $K$. 
We highlight that this exponential scaling in $K$ is due to the growth of the output calculated by our algorithm, i.e. the number of causal paths of length up to $K$ that can exist in the data.

\section{Discussion and Outlook}
\label{Discussion}

Causal paths are crucial to understand how nodes in systems with time-varying topologies indirectly influence each other.
This has recently been identified as an important problem in the study of time series data on complex networks~\cite{Lambiotte2019_Understanding}.
Fast methods to count causal paths are needed to fit and select higher-order models, detect paths with anomalous frequencies, model diffusion and epidemic spreading, rank nodes based on centralities, and detect clusters in big time series data on networks.
To this end, we present a fast streaming algorithm to count causal paths in temporal networks.
The results of our preliminary experiments (i) confirm our theoretical analysis of scalability, and (ii) show that it outperforms a baseline algorithm implemented in an OpenSource package.
Since both the baseline method and our proposed algorithm are deterministic, our experimental results are based on a single run in a single data set.
More experiments are thus needed to substantiate our findings in a larger number of big time series data.
Our method further has considerable parallelisation potential, which we will explore in future work.
In the light of these perspective, our work is a promising first step towards scalable data mining techniques for temporal network data.

\bibliographystyle{naturemag}
\bibliography{refs}

\begin{thebibliography}{10}
\expandafter\ifx\csname url\endcsname\relax
  \def\url#1{\texttt{#1}}\fi
\expandafter\ifx\csname urlprefix\endcsname\relax\def\urlprefix{URL }\fi
\providecommand{\bibinfo}[2]{#2}
\providecommand{\eprint}[2][]{\url{#2}}

\bibitem{page1999PageRank}
\bibinfo{author}{Page, L.}, \bibinfo{author}{Brin, S.},
  \bibinfo{author}{Motwani, R.} \& \bibinfo{author}{Winograd, T.}
\newblock \bibinfo{title}{The pagerank citation ranking: Bringing order to the
  web.}
\newblock \bibinfo{type}{Technical Report} \bibinfo{number}{1999-66},
  \bibinfo{institution}{Stanford InfoLab} (\bibinfo{year}{1999}).
\newblock \urlprefix\url{http://ilpubs.stanford.edu:8090/422/}.

\bibitem{rosvall2008maps}
\bibinfo{author}{Rosvall, M.} \& \bibinfo{author}{Bergstrom, C.~T.}
\newblock \bibinfo{title}{Maps of random walks on complex networks reveal
  community structure}.
\newblock \emph{\bibinfo{journal}{Proceedings of the National Academy of
  Sciences}} \textbf{\bibinfo{volume}{105}}, \bibinfo{pages}{1118--1123}
  (\bibinfo{year}{2008}).

\bibitem{grover2016node2vec}
\bibinfo{author}{Grover, A.} \& \bibinfo{author}{Leskovec, J.}
\newblock \bibinfo{title}{node2vec: Scalable feature learning for networks}.
\newblock In \emph{\bibinfo{booktitle}{Proceedings of the 22nd ACM SIGKDD
  international conference on Knowledge discovery and data mining}},
  \bibinfo{pages}{855--864} (\bibinfo{organization}{ACM},
  \bibinfo{year}{2016}).

\bibitem{holme2015modern}
\bibinfo{author}{Holme, P.}
\newblock \bibinfo{title}{Modern temporal network theory: a colloquium}.
\newblock \emph{\bibinfo{journal}{The European Physical Journal B}}
  \textbf{\bibinfo{volume}{88}}, \bibinfo{pages}{234} (\bibinfo{year}{2015}).

\bibitem{Lambiotte2019_Understanding}
\bibinfo{author}{{Lambiotte}, R.}, \bibinfo{author}{{Rosvall}, M.} \&
  \bibinfo{author}{{Scholtes}, I.}
\newblock \bibinfo{title}{{From networks to optimal higher-order models of
  complex systems}}.
\newblock \emph{\bibinfo{journal}{Nature Physics}}
  \textbf{\bibinfo{volume}{15}} (\bibinfo{year}{2019}).

\bibitem{scholtes2014causality}
\bibinfo{author}{Scholtes, I.} \emph{et~al.}
\newblock \bibinfo{title}{Causality-driven slow-down and speed-up of diffusion
  in non-markovian temporal networks}.
\newblock \emph{\bibinfo{journal}{Nat. Comm.}} \textbf{\bibinfo{volume}{5}},
  \bibinfo{pages}{5024} (\bibinfo{year}{2014}).

\bibitem{Rosvall2014_Memory}
\bibinfo{author}{Rosvall, M.}, \bibinfo{author}{Esquivel, A.~V.},
  \bibinfo{author}{Lancichinetti, A.}, \bibinfo{author}{West, J.~D.} \&
  \bibinfo{author}{Lambiotte, R.}
\newblock \bibinfo{title}{Memory in network flows and its effects on spreading
  dynamics and community detection}.
\newblock \emph{\bibinfo{journal}{Nature communications}}
  \textbf{\bibinfo{volume}{5}}, \bibinfo{pages}{4630} (\bibinfo{year}{2014}).

\bibitem{xu2016representing}
\bibinfo{author}{Xu, J.}, \bibinfo{author}{Wickramarathne, T.~L.} \&
  \bibinfo{author}{Chawla, N.~V.}
\newblock \bibinfo{title}{Representing higher-order dependencies in networks}.
\newblock \emph{\bibinfo{journal}{Science advances}}
  \textbf{\bibinfo{volume}{2}}, \bibinfo{pages}{e1600028}
  (\bibinfo{year}{2016}).

\bibitem{caceres2013temporal}
\bibinfo{author}{Caceres, R.~S.} \& \bibinfo{author}{Berger-Wolf, T.}
\newblock \bibinfo{title}{Temporal scale of dynamic networks}.
\newblock In \emph{\bibinfo{booktitle}{Temporal Networks}},
  \bibinfo{pages}{65--94} (\bibinfo{publisher}{Springer},
  \bibinfo{year}{2013}).

\bibitem{Kempe2000}
\bibinfo{author}{Kempe, D.}, \bibinfo{author}{Kleinberg, J.} \&
  \bibinfo{author}{Kumar, A.}
\newblock \bibinfo{title}{Connectivity and inference problems for temporal
  networks}.
\newblock In \emph{\bibinfo{booktitle}{Proceedings of the thirty-second annual
  ACM symposium on Theory of computing}}, \bibinfo{pages}{504--513}
  (\bibinfo{organization}{ACM}, \bibinfo{year}{2000}).

\bibitem{wu2014path}
\bibinfo{author}{Wu, H.} \emph{et~al.}
\newblock \bibinfo{title}{Path problems in temporal graphs}.
\newblock \emph{\bibinfo{journal}{Proceedings of the VLDB Endowment}}
  (\bibinfo{year}{2014}).

\bibitem{Whitbeck2012}
\bibinfo{author}{Whitbeck, J.}, \bibinfo{author}{Dias~de Amorim, M.},
  \bibinfo{author}{Conan, V.} \& \bibinfo{author}{Guillaume, J.-L.}
\newblock \bibinfo{title}{Temporal reachability graphs}.
\newblock In \emph{\bibinfo{booktitle}{Proceedings of the 18th Annual
  International Conference on Mobile Computing and Networking}}, Mobicom '12
  (\bibinfo{year}{2012}).

\bibitem{Lentz2013}
\bibinfo{author}{Lentz, H. H.~K.}, \bibinfo{author}{Selhorst, T.} \&
  \bibinfo{author}{Sokolov, I.~M.}
\newblock \bibinfo{title}{Unfolding accessibility provides a macroscopic
  approach to temporal networks}.
\newblock \emph{\bibinfo{journal}{Phys. Rev. Lett.}}
  \textbf{\bibinfo{volume}{110}}, \bibinfo{pages}{118701}
  (\bibinfo{year}{2013}).

\bibitem{Takaguchi2016_CoverageCentrality}
\bibinfo{author}{Takaguchi, T.}, \bibinfo{author}{Yano, Y.} \&
  \bibinfo{author}{Yoshida, Y.}
\newblock \bibinfo{title}{Coverage centralities for temporal networks}.
\newblock \emph{\bibinfo{journal}{The European Physical Journal B}}
  \textbf{\bibinfo{volume}{89}}, \bibinfo{pages}{35} (\bibinfo{year}{2016}).

\bibitem{salnikov2016using}
\bibinfo{author}{Salnikov, V.}, \bibinfo{author}{Schaub, M.~T.} \&
  \bibinfo{author}{Lambiotte, R.}
\newblock \bibinfo{title}{Using higher-order markov models to reveal flow-based
  communities in networks}.
\newblock \emph{\bibinfo{journal}{Scientific reports}}
  \textbf{\bibinfo{volume}{6}}, \bibinfo{pages}{23194} (\bibinfo{year}{2016}).

\bibitem{scholtes2017network}
\bibinfo{author}{Scholtes, I.}
\newblock \bibinfo{title}{When is a network a network?: Multi-order graphical
  model selection in pathways and temporal networks}.
\newblock In \emph{\bibinfo{booktitle}{Proceedings of the 23rd ACM SIGKDD
  International Conference on Knowledge Discovery and Data Mining}}
  (\bibinfo{year}{2017}).

\bibitem{paranjape2017motifs}
\bibinfo{author}{Paranjape, A.}, \bibinfo{author}{Benson, A.~R.} \&
  \bibinfo{author}{Leskovec, J.}
\newblock \bibinfo{title}{Motifs in temporal networks}.
\newblock In \emph{\bibinfo{booktitle}{Proceedings of the Tenth ACM
  International Conference on Web Search and Data Mining}},
  \bibinfo{pages}{601--610} (\bibinfo{organization}{ACM},
  \bibinfo{year}{2017}).

\bibitem{Eagle2006}
\bibinfo{author}{Eagle, N.} \& \bibinfo{author}{(Sandy)~Pentland, A.}
\newblock \bibinfo{title}{Reality mining: sensing complex social systems}.
\newblock \emph{\bibinfo{journal}{Personal and Ubiquitous Computing}}
  \textbf{\bibinfo{volume}{10}}, \bibinfo{pages}{255--268}
  (\bibinfo{year}{2006}).

\end{thebibliography}

\end{document}